\documentclass[final,nohyperref]{dmtcs-episciences}

\usepackage[utf8]{inputenc}
\usepackage{subfigure}
\usepackage{graphicx}
\usepackage[round]{natbib}
\usepackage{amsmath,amssymb,amsthm}

\newcommand{\dir}[1]{\overrightarrow{#1}}

\newcommand{\dprime}{{\prime\prime}}

\newtheorem{theorem}{Theorem}[section]
\newtheorem{lemma}[theorem]{Lemma}
\newtheorem{corollary}[theorem]{Corollary}

\author{Christopher Duffy\affiliationmark{1}\thanks{This author acknowledges the support of the Natural Sciences and Engineering Research Council of Canada (NSERC)}
	\and Sonja Linghui Shan\affiliationmark{2}}

\title[Homomorphisms of oriented and \(2\)-edge-coloured graphs to reflexive targets]{On the existence and non-existence of improper homomorphisms of oriented and \(2\)-edge-coloured graphs to reflexive targets}

\affiliation{Department of Mathematics and Statistics, University of Saskatchewan, Canada\\
	Department of Computer Science, University of Saskatchewan, Canada}


\received{2020-09-11}
\revised{2021-02-17}
\accepted{2021-02-26}

\begin{document}
	\publicationdetails{23}{2021}{1}{6}{6773}
	\maketitle
	\begin{abstract}
		We consider non-trivial homomorphisms to reflexive oriented graphs in which some pair of adjacent vertices have the same image.
		By way of a notion of convexity for oriented graphs, we study those oriented graphs that do not admit such  homomorphisms.
		We fully classify those oriented graphs with tree-width \(2\) that do not admit such  homomorphisms and show that it is NP-complete to decide if a  graph admits an orientation that does not admit such  homomorphisms.
		We prove  analogous results for \(2\)-edge-coloured graphs.
		We apply our results on oriented graphs to provide a new tool in the study of chromatic number of orientations of planar graphs -- a long-standing open problem.
	\end{abstract}

\section{Introduction and Preliminaries}
The main matter of this paper concerns homomorphisms of two objects that arise from graphs with no parallel edges -- oriented graphs and \(2\)-edge-coloured graphs.
An oriented graph, \(\dir{G}\), arises from a graph \(G\) by assigning to each edge a direction to form an arc.
Alternately an oriented graph is an anti-symmetric digraph.
A \(2\)-edge-coloured graph arises from a graph \(G\) by assigning a colour, red or blue, to each edge.
Alternately, a \(2\)-edge-coloured graph is a pair \((G,c_G)\) where \(G = (V_G,E_G)\) is a graph and \(c_G:E_G \to \{R,B\}\).
In each case we refer to \(G\) as the \emph{underlying graph}.
As various portions of this work deals with reflexive and irreflexive graphs, oriented graphs and \(2\)-edge-coloured graphs, we permit these objects to have loops unless otherwise specified.
Reflexive \(2\)-edge-coloured graphs are assumed to have a loop of each edge colour at every vertex.
To simplify matters, we assume graphs herein are connected and have at least two vertices unless otherwise specified.
For other notation and definitions not defined herein we refer to \cite{BM08}.

For an oriented graph \(\dir{G}\) with \(uv, vw \in A_{\dir{G}}\) with \(u\neq v\) and \(v\neq w\) we say that \(u,v,w\) is a \(2\)-dipath and that \(v\) is the \emph{centre} of the \(2\)-dipath.
For a \(2\)-edge-coloured graph \((G,c_G)\) with \(uv, vw \in E_{G}\) with \(u\neq v\) and \(v\neq w\) we say that \(u,v,w\) is a \(2\)-path, and that \(v\) is the \emph{centre} of the \(2\)-path.
When \(c_G(uv) \neq c_G(vw)\) we say  the \(2\)-path is \emph{alternating}.
Otherwise we say the \(2\)-path is \emph{monochromatic}.

Let \(\dir{G}\) and \(\dir{H}\) be  oriented graphs.
We say \emph{there is a homomorphism of \(\dir{G}\) to \(\dir{H}\)} when there exists \(\phi:V_{G} \to V_{H}\) so that for each \(uv \in A_{\dir{G}}\) we have \(\phi(u)\phi(v) \in A_{\dir{H}}\).
When there is a homomorphism of \(\dir{G}\) to \(\dir{H}\) we write \({\dir{G}} \to \dir{H}\). 
We say \(\phi\) is a \emph{homomorphism} and we write \(\phi: {\dir{G}} \to \dir{H}\).
Notice that \(\phi\) induces a mapping from the arc set of \(\dir{G}\) to the arc set of \(\dir{H}\).

The study of oriented graph, and more generally, directed graph and graph homomorphisms has a rich history.
Countless of pages of research articles, monographs and theses have been devoted to their study.
Some of this focus comes by way of constraint satisfaction problems, which can be re-framed as questions of existence of certain directed graph homomorphisms.
And so the computational complexity the \(\dir{H}\)-colouring problem has been of particular interest.
\cite{B17} and \cite{Z17} independently verified the CSP Dichotomy Conjecture (see \cite{F93}).
This result implies the \(\dir{H}\)-colouring problem can be classified as Polynomial or NP-complete based on the existence of a near-weak unanimity function for \(\dir{H}\).

Another major research area resulting from oriented graph homomorphisms comes by way of the generalization of graph colouring to irreflexive oriented graphs.
\cite{CO94} first introduced these colourings as an example of a graph property expressible in the monadic second order logic of graphs.
We return to this topic in Section \ref{sec:MinorClosed} and so eschew definitions and further details here.
We, however, point the eager reader to the survey by \cite{S16}. 

For any oriented graphs \({\dir{G}}\) and \(\dir{H}\) it is always true  \({\dir{G}} \to \dir{H}\) provided \(\dir{H}\) is reflexive.
Indeed, one may consider the trivial homomorphism that maps each vertex of \(\dir{G}\) to the same vertex of \(\dir{H}\).
Here the induced mapping on the arc sets maps every arc of \(\dir{G}\) to the same loop in \(\dir{H}\).
Of course for  some choices of \(\dir{G}\) and \(\dir{H}\) there exist non-trivial homomorphisms and even homomorphisms for which the induced mapping on the arc set uses no loop.
For example, one can  verify that there is a homomorphism of any orientation of a tree to the reflexive directed \(3\)-cycle in which the induced mapping on the arc set uses no loop in the reflexive \(3\)-cycle.

Let \(\dir{G}\) be an oriented graph and let \(\dir{H}\) be a reflexive oriented graph.
Let \(\phi: \dir{G} \to \dir{H}\) be a homomorphism.
The homomorphism \(\phi\) can be classified as one of three types.
We say \(\phi\) is \emph{trivial} when the induced mapping from the arc set of \(\dir{G}\) to that of \(\dir{H}\) maps every arc to the same loop.
We say \(\phi\) is \emph{improper} when \(\phi\) is not trivial and the induced mapping from the arc set of \(\dir{G}\) to that of \(\dir{H}\) maps at least one arc to a loop.
Finally we say \(\phi\) is \emph{proper} when the induced mapping from the arc set of \(\dir{G}\) to that of \(\dir{H}\) maps no arc to a loop.

Our analysis of homomorphisms to reflexive oriented graphs remain the same when we replace oriented graph with \(2\)-edge-coloured graph. 
In this case, homomorphism requires that existence and colour of edges are preserved.
And so we define \emph{trivial, improper} and \emph{proper} analogously for homomorphisms \(2\)-edge-coloured graphs.

The main aim of this work is to study the structure of those oriented graphs and \(2\)-edge-coloured graphs that admit no improper homomorphisms.
To aid our study we consider the following notion of convexity for oriented and \(2\)-edge-coloured graphs.

Let \(\dir{G}\) be an oriented graph and consider \(S \subseteq V_{\dir{G}}\).
We say \(S\) is \emph{convex} when no vertex in \(V_{\dir{G}} \setminus S\) is the centre vertex of a \(2\)-dipath whose ends are in \(S\).
The \emph{convex hull of \(S\)}, denoted \(conv(S)\), is the smallest subset of \(V_G\) so that \(S \subseteq conv(S)\) and \(conv(S)\) is convex.
When \(\dir{H}\) is an induced subgraph of \(\dir{G}\) we use \(conv(\dir{H})\) to denote \(conv(V_{\dir{H}})\).

We analogously define \emph{convex} and \emph{convex hull} for \(2\)-edge-coloured graphs by replacing \(2\)-dipath with alternating \(2\)-path in our definition.

Equivalently one may define the convex hull of a set of vertices as follows.
Let \(\Gamma\) be an oriented or \(2\)-edge-coloured graph and consider \(S \subseteq V_\Gamma\).
Define the sequence \(S_0, S_1, \dots\) so that
\begin{itemize}
	\item \(S_0 = S\); and
	\item for each \(1 \leq i \leq k\) let \(S_{i+1} = S_i \cup N\), where \(N\) is the set of vertices that are the centre of a \(2\)-dipath or alternating \(2\)-path whose ends are in \(S_i\).
\end{itemize}

Since \(S_0 \subseteq S_1 \subseteq \dots\) and \(V_\Gamma\) is finite, there exists a least integer \(k\) so that \(S_{k}=S_{k+1}\).
IOne may verify  \(S_k = conv(S)\). 
This in turn implies that our notion of convex hull is well-defined.

\begin{figure}
	\begin{center}
		\includegraphics[width=0.5\linewidth]{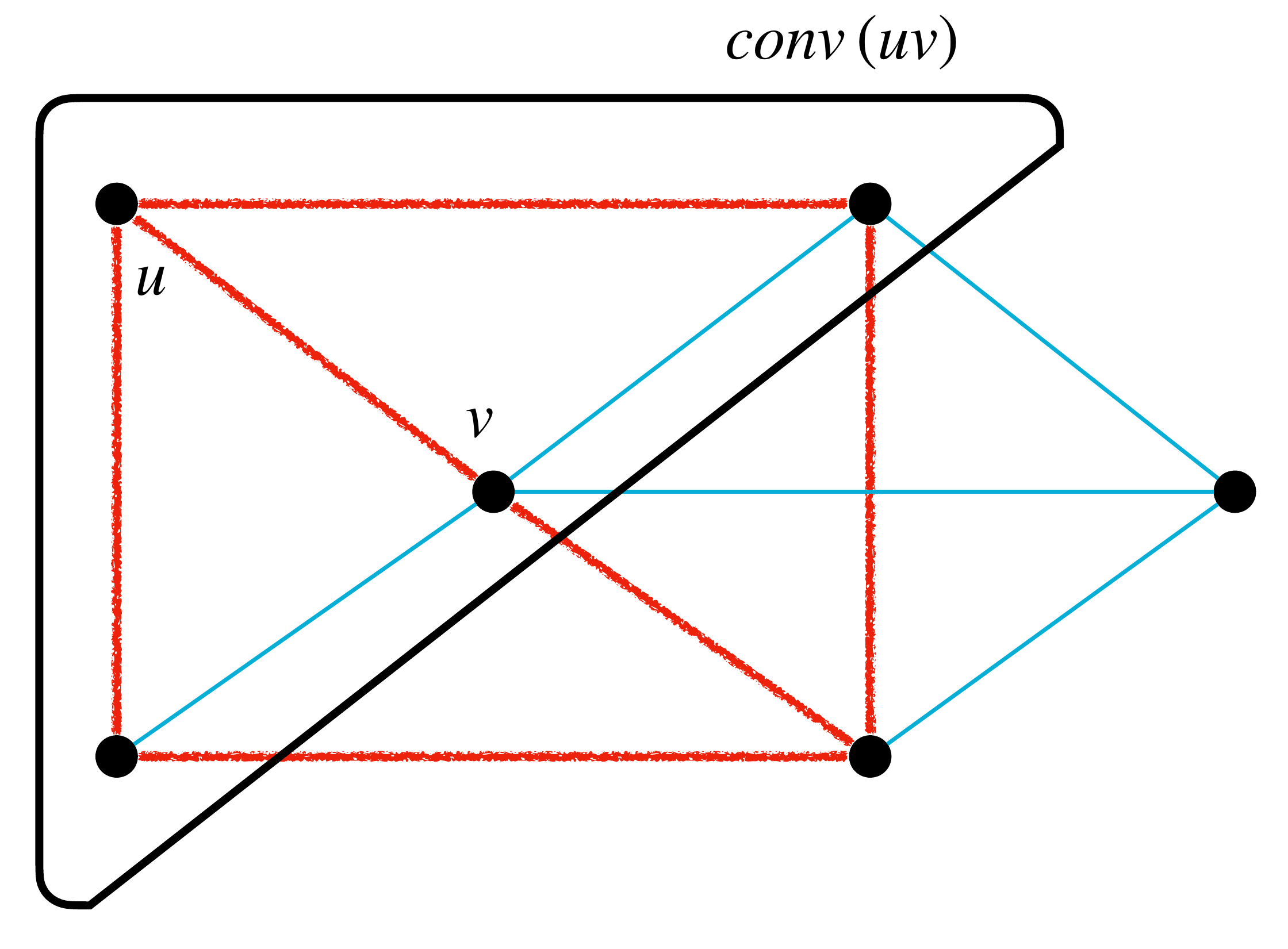}
		\caption{The convex hull of \(u\) and \(v\)}
		\label{fig:convHullExample}
	\end{center}
\end{figure}

With this definition it is clear that  if \(S^\prime\subseteq S\), then \(conv(S^\prime) \subseteq conv(S)\).
By way of example, consider the \(2\)-edge-coloured graph in Figure \ref{fig:convHullExample}.

\cite{S02b} introduces this notion of convexity for oriented graphs.
\cite{DP18} extend to notion \(2\)-edge-coloured graphs.
In both cases the following key observation arises.

\begin{lemma}\label{lem:sameCol} \cite{DP18,S02b}
	Let \(\Gamma\) be an oriented  or \(2\)-edge-coloured graph. 
	Let \(\phi\) be a homomorphism of \(\Gamma\) to a reflexive target. 
	If for \(uv \in A_\Gamma\) ($uv \in E_\Gamma$) we have \(\phi(u) = \phi(v)\), then \(\phi(x) = \phi(u)\) for each \(x \in conv(uv)\).
\end{lemma}

\begin{figure}
\begin{center}
		\includegraphics[width=0.5\linewidth]{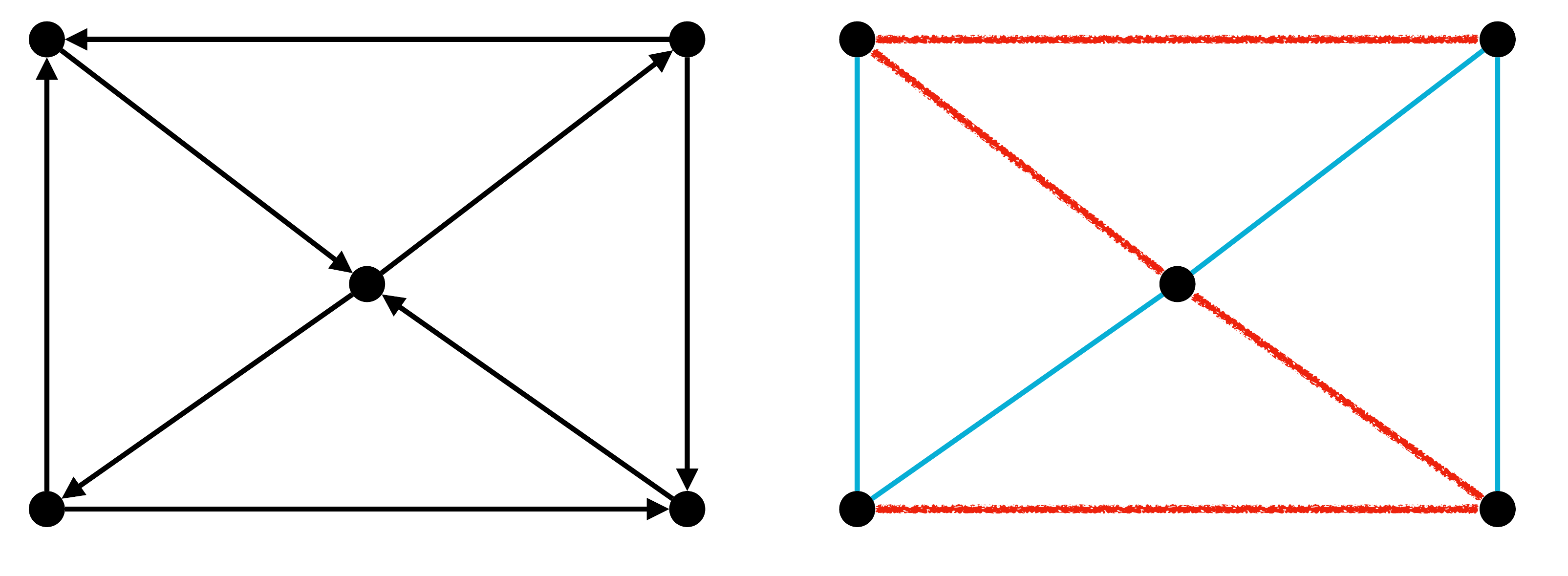}
\end{center}
	\caption{A complete-convex oriented graph and a complete-convex \(2\)-edge-coloured graph}
	\label{fig:OrientCompleteConvexExample}
\end{figure}

Let \(\Gamma\) be an oriented or \(2\)-edge-coloured graph.
We say that \(\Gamma\) is \emph{complete convex} when for each \(uv \in A_{\dir{\Gamma}}\) (\(uv \in E_\Gamma\)) we have \(conv(uv) = V_{\Gamma}\).
Notice that the \(2\)-edge-coloured graph in Figure \ref{fig:convHullExample} is not complete convex.
See Figure \ref{fig:OrientCompleteConvexExample} for an example of a complete-convex oriented graph and complete-convex \(2\)-edge-coloured graph.

\begin{theorem}\label{thm:onlyG}
	Let \(\Gamma\) be an oriented or \(2\)-edge-coloured graph. We have that \(\Gamma\) admits no improper homomorphism if and only if \(\Gamma\) is  complete convex. 
\end{theorem}

We prove the result for oriented graphs.
The result for \(2\)-edge-coloured graphs follows similarly.

\begin{proof}
	Let \(\dir{G}\) be a complete-convex oriented graph and let \(\dir{H}\) be a reflexive oriented graph.
	If there exists \(\phi: \dir{G} \to \dir{H}\) that is improper, then there exists \(uv \in A_G\) so that \(\phi(u) = \phi(v)\).
	By Lemma \ref{lem:sameCol} we have \(\phi(x) = \phi(u)\) for each \(x \in conv(uv)\).
	Since \(\dir{G}\) is complete convex it we have \(conv(uv) = V_G\).
	Therefore \(\phi\) is trivial, a contradiction.
	
	Let \(\dir{G}\) be an oriented graph so that for every reflexive oriented graph \(\dir{H}\) and every homomorphism \(\phi: \dir{G} \to \dir{H}\) we have that \(\phi\) is either trivial or proper.
	Let \(V_G = \{v_1,v_2,\dots, v_n\}\).
	
	If \(\dir{G}\) is not complete convex then there exists \(uv \in A_G\) so that \(conv(uv) \neq V_G\).
	Let \(S = conv(uv)\).
	Let \(k = |S|\).
	Without loss of generality, let \(S = \{v_1,v_2, \dots v_{k}\}\).
	Consider \(c: V_G \to \{k, k+1, k+2,\dots, n\}\) so that \(c(v_t) = t\) for all \(t > k\) and \(c(v_t) = k\) for all \(v_t \in S\).
	The mapping \(c\) defines a improper homomorphism to a reflexive oriented graph \(\dir{H}\) with vertex set \(\{k,k+1,k+2,\dots,n\}\) where \(ij \in A_H\) whenever \(v_iv_j \in A_G\) ($k \leq i < j \leq n$).
	This contradicts the non-existence of improper homomorphisms of \(\dir{G}\).
\end{proof}

Theorem \ref{thm:onlyG} implies that we may study oriented or \(2\)-edge-coloured graphs that admit no improper homomorphism using the notion of convexity introduced above.
Since one may check in polynomial time whether an oriented or \(2\)-edge-coloured graph is complete convex, we arrive at the following.

\begin{theorem}
	For any pair of input  oriented or \(2\)-edge-coloured graphs \(\Gamma_1\) and \(\Gamma_2\) with \(\Gamma_2\) reflexive, it is Polynomial to decide if the only homomorphisms of \(\Gamma_1\) to \(\Gamma_2\) are each either trivial or proper.
\end{theorem}

The remainder of the work proceeds as follows:
In Section \ref{sec:Structure} we study the structure of oriented and \(2\)-edge-coloured graphs that admit no improper homomorphisms.
In doing so we fully classify such oriented  and \(2\)-edge-coloured graphs whose underlying graphs have tree-width \(2\). 
In Section \ref{sec:Complexity}, we show to be NP-complete  the problem of deciding if a graph is the underlying graph of some \(2\)-edge-coloured or oriented graph that admits no improper homomorphism. 
In Section \ref{sec:MinorClosed} we apply our work on improper homomorphisms to the study of the chromatic number of oriented graphs.
In doing so we provide a tool that may aid work on a long-standing open-problem concerning the chromatic number of oriented planar graphs.
We discuss this and other impacts of our work in Section \ref{sec:Conclusion}.

\section{The Structure of Oriented and \(2\)-edge Coloured Graphs that Admit No Improper Homomorphism}\label{sec:Structure}

We apply Theorem \ref{thm:onlyG} and study complete convexity in oriented and \(2\)-edge-coloured graphs.
We begin with two n\"{a}ive, but useful, observations.

\begin{lemma}\label{lem:OrientID}
	Let \(\dir{G}\) and \(\dir{H}\) be oriented graphs. If \(\dir{G}\) and \(\dir{H}\) are each complete convex, then any oriented graph formed by identifying any arc of \(\dir{G}\) with any arc of \(\dir{H}\) is complete convex.
\end{lemma}

\begin{proof}
	Let \(\dir{G}\) and \(\dir{H}\) be complete-convex oriented graphs.
	Consider \(g \in A_{\dir{G}}\) and \(h \in A_{\dir{H}}\).
	Let \(\dir{J}\) be the oriented graph produced by identifying \(g\) and \(h\).
	Let \(g = h = xy\).
	We observe \(conv_{\dir{J}}(xy) = V_G \cup V_H\).
	Since \(\dir{G}\) and \(\dir{H}\) are each complete convex, for any \(uv \in A_{\dir{G}} \cup A_{\dir{H}}\)  we have \(x,y \in conv_{\dir{J}}(uv)\).
	Therefore \(conv_{\dir{J}}(uv) = V_G \cup V_H\).
	Therefore \(\dir{J}\) is complete convex.
\end{proof}

A similar result holds for \(2\)-edge-coloured graphs.

\begin{lemma}\label{lem:2ecId}
	Let \((G,c_G)\) and \((H,c_H)\) be \(2\)-edge-coloured graphs. If \((G,c_G)\) and \((H,c_H)\) are complete convex, then any \(2\)-edge-coloured graph formed from by identifying a red (blue) edge in \((G,c_G)\) and a red (blue) edge in  \((H,c_H)\) is complete convex.
\end{lemma}

With an eye towards graphs with tree-width \(2\), we continue our study with an observation of vertices of degree \(2\) in complete-convex oriented  graphs.

\begin{lemma}\label{lem:degree2Triangle}
	Let \(G\) be a graph and let \(v\) be a vertex of degree \(2\) in \(G\).
	If \(\dir{G}\) is a complete-convex oriented graph, then \(v\) is contained in a directed \(3\)-cycle in \(\dir{G}\).
\end{lemma}

\begin{proof}
	Let \(\dir{G}\) be a complete-convex oriented graph and \(v\) be a vertex with degree \(2\) in \(G\).
	Let \(x\) and \(y\) be the neighbours of \(v\) in \(G\).
	Since \(\dir{G}\) is complete convex, each vertex of \(\dir{G}\) must be the centre vertex of a \(2\)-dipath in \(\dir{G}\).
	Without loss of generality, assume \(xv, vy \in A_{\dir{G}}\).
	If \(yx \notin A_{\dir{G}}\), then \(conv(xv) = \{x,v\} \neq V_{\dir{G}}\).
	Thus the subgraph induced by \(\{x,y,z\}\) is a directed \(3\)-cycle.	
\end{proof}

\begin{lemma}\label{lem:degree2Reduce}
	Let \(\dir{G}\) be an oriented graph with a vertex \(v\) of degree \(2\).
	If \(\dir{G}\) is complete convex, then \(\dir{G-v}\) is complete convex.
\end{lemma}

\begin{proof}
	Assume \(\dir{G}\) is complete convex.  
	Consider the oriented graph \(\dir{G-v}\).
	If \(\dir{G-v}\) is not complete convex, then there exists \(uw \in A_{\dir{G-v}}\) so that \(conv_{\dir{G-v}}(uw) \neq V_{G-v}\).
	Let \(S = conv_{\dir{G-v}}(uw)\).
	Since \(\dir{G}\) is complete convex, it must be that \(conv_{\dir{G}}(S)= V_G\).
	Since \(S \neq V_{G-v}\), the set \(V_{G-v} \setminus S\) is non-empty.
	Since \(conv_{\dir{G}}(S) \neq S\), there exists \(q \in V_{G}\setminus S\) so that \(q\) is the centre vertex of a \(2\)-dipath whose ends are in \(S\).
	This \(2\)-dipath does not exist in \(\dir{G-v}\), thus \(q = v\) and so \(x,y \in S\).
	
	Consider now \(S \cup v\).
	Again, since \(V_{G-v} \setminus S\) is non-empty and \(conv_{\dir{G}}(S \cup v) = V_G\), there exists \(q^\prime \in V_{G} \setminus (S\cup v)\) so that \(q^\prime\) is the centre vertex of a \(2\)-dipath whose ends are in \(S\cup v\).
	Since \(x,y \in S\),  vertex \(v\) cannot be an end of this \(2\)-dipath.
	Let \(s_1,s_2\) be the ends of this \(2\)-dipath.
	Notice that \(V_{G} \setminus (S\cup v) = V_{G-v} \setminus S\).
	Thus \(s_1q^\prime s_2\) is a \(2\)-dipath whose ends are in \(S\) and whose centre vertex is in  \(V_{G-v} \setminus S\).
	This contradicts that \(conv_{\dir{G-v}}(S) = S\).
	Therefore \(\dir{G-v}\) is  complete convex.
\end{proof}

Lemmas \ref{lem:degree2Triangle} and \ref{lem:degree2Reduce} allow us to fully classify those graphs with tree-width \(2\) that admit a complete-convex orientation.

\begin{theorem}\label{thm:2treeOrient}
	Let \(T\) be a graph with tree-width  \(2\).
	An orientation of \(T\),  say \(\dir{T}\), is complete convex if and only if \(T\) is a \(2\)-tree and every induced copy of \(K_3\) in \(T\) is a directed \(3\)-cycle in \(\dir{T}\).
\end{theorem}

\begin{proof}
	Assume \(T\) is a \(2\)-tree.
	Let \(\dir{T}\) be an orientation of \(T\) in which every induced \(K_3\) is oriented as a directed \(3\)-cycle.
	We show \(\dir{T}\) is complete convex by induction on \(n\), the number of vertices of \(T\).
	Notice that the directed \(3\)-cycle is complete convex.
	Assume that \(\dir{T}\) has \(k+1 > 3\) vertices.
	Let  \(v\) be a vertex of degree \(2\).
	By induction \(\dir{T-v}\) is complete convex.
	It follows from Lemma \ref{lem:OrientID} that \(\dir{T}\) is complete convex.
	
	Consider now a graph \(T\) with tree-width \(2\) so that  \(\dir{T}\) is complete convex.
	Since \(T\) has tree-width \(2\), \(T\) is a spanning subgraph of a \(2\)-tree, \(T^\prime\).
	Since \(T^\prime\) is a \(2\)-tree, there is an ordering of its vertices: \(v_1,v_2,\dots, v_n\) so that \(T^\prime[\{v_1,v_2\}] = K_2\) and for each \(1 \leq i \leq n\) vertex \(v_i\) has degree \(2\) and is contained in a copy of \(K_3\) in \(T^\prime[\{v_1,v_2,\dots, v_i\}]\).
	
	If \(T\) is a proper subgraph of \(T^\prime\), then there is a largest index \(j\) such that \(v_j\) has degree less than \(2\) in \(T^\prime[\{v_1,v_2,\dots, v_j\}]\).
	By Lemma \ref{lem:degree2Reduce}, the oriented graph \(\dir{T}[\{v_1,v_2,\dots, v_j\}]\) is complete convex.
	The only complete-convex oriented graph with a vertex of degree \(1\) is \(\dir{K_2}\).
	Therefore  \(\dir{T}[\{v_1,v_2,\dots, v_j\}]\) is not complete convex.
	This is a contradiction.
	Thus \(T\) is a \(2\)-tree.
	
	It remains to show that every induced copy of \(K_3\) is a directed \(3\)-cycle in \(\dir{T}\).
	Assume otherwise.
	Thus there exists a greatest index \(k>2\) so that the copy of \(K_3\) in \(T[\{v_1,v_2,\dots, v_k\}]\) that contains \(v_k\) is not a directed \(3\)-cycle in \(\dir{T}[\{v_1,v_2,\dots, v_k\}]\).
	By Lemma \ref{lem:degree2Triangle}, \(\dir{T}[\{v_1,v_2,\dots, v_k\}]\) is complete convex.
	This contradicts the statement of Lemma  \ref{lem:degree2Triangle} as \(v_k\) has degree \(2\) in \(T[\{v_1,v_2,\dots, v_k\}]\).
\end{proof}

Using Theorem \ref{thm:onlyG} we  re-frame these results to apply to the study of oriented graphs that admit no improper homomorphism.

\begin{corollary}\label{cor:2treeorient}
	An orientation of a graph with tree width \(2\), \(\dir{T}\), admits no improper homomorphism if and only if \(T\) is a \(2\)-tree and every induced copy of \(K_3\) in \(T\) is a directed \(3\)-cycle in \(\dir{T}\).
\end{corollary}

Lemma \ref{lem:degree2Reduce} gives a method to add/remove vertices of degree \(2\) to a complete-convex oriented graph so that the resulting oriented graph is complete convex:
Let \(\dir{H}\) be a complete-convex oriented graph with \(k\geq 2\) vertices.
For any arc \(uw \in A_{\dir{H}}\) one may add a new vertex \(v\) and arcs \(wv,vu\) to form a complete-convex oriented graph with \(k+1\) vertices.
We now describe a method of adding/removing arcs to a complete-convex oriented graph so that the resulting oriented graph is complete convex.

Let \(\dir{G}\) be oriented graph and let \(a \in A_{\dir{G}}\).
Denote by \(\dir{G_a}\) the oriented graph formed from \(\dir{G}\) by reversing the orientation of \(a\).
Denote by \(\dir{G-a}\) the oriented graph formed from \(\dir{G}\) by removing \(a\).

\begin{theorem}\label{thm:ReverseArc}
	Let \(\dir{G}\) be an oriented graph and let \(a \in A_{\dir{G}}\).
	If \(\dir{G}\) and \(\dir{G_a}\) are complete convex, then \(\dir{G-a}\) is complete convex.
\end{theorem}

\begin{proof}
	Let \(\dir{G}\) be an oriented graph and let \(uv \in A_{\dir{G}}\) so that \(\dir{G}\) and \(\dir{G_{uv}}\) are complete convex.
	If 	\(\dir{G-uv}\) is not complete convex then there exists \(xy \in A_{\dir{G-uv}}\) so that \(conv_{\dir{G-uv}}(xy) \neq V_{G}\).
	Let \(S = conv_{\dir{G-uv}}(xy)\).
	If \(u,v \in S\) or \(u,v \notin S\), then \(conv_{\dir{G-uv}}(xy) =conv_{\dir{G}}(xy)  = V_G\).
	Thus, without loss of generality, assume \(u \in S\) and \(v \notin S\).
	Since \(conv_{\dir{G}}(S) = V_G\), but \(conv_{\dir{G-uv}}(S) = S\), there exists \(w \in S\) so that \(vw \in A_{\dir{G}}\).
	Similarly, since \(conv_{\dir{G_{uv}}}(S) = V_G\), but \(conv_{\dir{G-uv}}(S) = S\), there exists \(w^\prime \in S\) so that \(w^\prime v \in A_{\dir{G}}\).
	Therefore \(w^\prime, v, w\) is a \(2\)-dipath \(\dir{G-uv}\) so that \(w,w^\prime \in S\) and \(v \notin S\).
	Therefore \(conv_{\dir{G-uv}}(S) \neq S\), a contradiction.
\end{proof}

\begin{theorem}\label{thm:AddArc}
	Let \(\dir{G}\) be a complete-convex oriented graph.
	If \(uv \notin E_G\) and \(u\) and \(v\) are the ends of a \(2\)-dipath in \(\dir{G}\), then each of \(\dir{G + uv}\) and \(\dir{G + vu}\) is complete convex.
\end{theorem}

\begin{proof}
	Assume \(\dir{G}\) is a complete-convex oriented graph.
	Consider  \(uv \notin E_G\) so that \(u,x,v\) is a \(2\)-dipath in \(\dir{G}\) for some \(x \in V_G\).
	Notice \(x \in conv_{\dir{G}}(\{u,v\})\).
	Therefore \(conv_{\dir{G}}(\{u,v\}) = conv_{\dir{G}}(\{u,v,x\})\).
	Since \(ux \in A_{\dir{G}}\), it follows that \(conv_{\dir{G}}(\{u,v\}) = conv_{\dir{G}}(\{u,v,x\}) = V_G\).
	Therefore \(conv_{\dir{G+uv}}(uv) = conv_{\dir{G}}(\{u,v\}) = V_G\).
	And so it follows that  \(\dir{G + uv}\) is complete convex.
	
	A similar argument shows \(\dir{G + vu}\) is complete convex.
\end{proof}

As with Theorem \ref{thm:2treeOrient}, each of Theorem \ref{thm:ReverseArc} and \ref{thm:AddArc} can be re-interpreted as statements about oriented graphs that admit no improper homomorphism.

\begin{corollary}
	Let \(\dir{G}\) be an oriented graph and let \(a \in A_{\dir{G}}\).
	If \(\dir{G}\) and \(\dir{G_a}\) each admit no improper homomorphism, then \(\dir{G-a}\) is admits no improper homomorphism.
\end{corollary}

\begin{corollary}
	Let \(\dir{G}\) be an oriented graph that admits no improper homomorphism.
	If \(uv \notin E_G\) and \(u\) and \(v\) are the ends of a \(2\)-dipath in \(\dir{G}\), then each of \(\dir{G + uv}\) and \(\dir{G + vu}\) admit no improper homomorphism.
\end{corollary}

For \(2\)-edge-coloured graphs a different picture emerges when we examine vertices of minimum degree.
\begin{theorem}
	A complete-convex \(2\)-edge-coloured graph is either a monochromatic copy of \(K_2\) or has minimum degree \(3\).
\end{theorem}

\begin{proof}
	Consider \(G \not\cong K_2\).
	Let \((G, c_G)\) be a complete-convex \(2\)-edge-coloured graph.
	Let \(v\) be a vertex of minimum degree in \(G\).
	
	If \(v\) has a single neighbour, say \(u\), in \(G\), then \(conv(uv) = \{u,v\} \neq V_G\).
	Assume \(v\) has two neighbours, say \(x\) and \(y\), in \(G\).
	Since \((G, c_G)\) is complete convex, \(v\) must be the centre vertex of an alternating \(2\)-path and \(x\) and \(y\) must be adjacent.
	Without loss of generality, let \(c_G{(xv)} = R, c_G{(vy)} =B\) and \(c_G{(xy)} = R\).
	However we notice that \(conv(vy) = \{v,y\} \neq V_G\).
\end{proof}

In fact, graphs that are not sufficiently dense cannot be the underlying graph of a complete-convex \(2\)-edge coloured graph.
\begin{theorem}\label{thm:2n-3}
	If \(G\) is a graph with at most  \(2n-3\) edges, then \(G\) is not the underlying graph of any complete-convex \(2\)-edge coloured graph.
\end{theorem}

\begin{proof}
	Let \(G\) be a graph with at most  \(2n-3\) edges. 
	If \(G\) has exactly two vertices, then \(G\) is a monochromatic copy of \(K_2\), which is not complete convex.
	
	Let \(G\) be a graph with \(n\geq 3\) vertices and at most \(2n-3\) edges. 
	Let \((G, c_G)\) be a \(2\)-edge-coloured graph.
	If \((G, c_G)\) is monochromatic then the convex hull of every edge contains only the end points of the edge.
	Therefore \((G, c_G)\) is not complete convex.
	
	Assume then that \((G, c_G)\) is not monochromatic.
	Without loss of generality, there are at most \(n-2\) red edges in \((G, c_G)\).
	Therefore the spanning subgraph formed by removing all blue edges is not connected.
	Let \(G_R\) be a non-trivial component of this subgraph.
	Let \(K_2^b\) denote the \(2\)-edge-coloured copy of \(K_2\) where the edge between the distinct vertices is blue.
	Let \(v_1\) and \(v_2\) be the vertices of \(K_2^b\).
	Consider the homomorphism \(\phi:(G, c_G) \to K_2^b\) given by
	\[\phi(x) = \begin{cases}
		v_1& x \in V(G_R)\\
		v_2 & \mbox{otherwise}
	\end{cases}\]
	Since \(G_R\) necessarily contains at least one edge, this homomorphism is improper.
	The result now follows by Theorem \ref{thm:onlyG}.
\end{proof}

\begin{corollary}
	No \(2\)-degenerate graph is the underlying graph of a complete-convex \(2\)-edge-coloured graph.
\end{corollary}

Contrasting these results with the statements of Theorem \ref{thm:2treeOrient} and Corollary \ref{cor:2treeorient}, we observe the following:
\begin{theorem}\label{thm:2tree2ec}
	No graph with tree-width \(2\) is the graph underlying a complete-convex \(2\)-edge-coloured graph.
\end{theorem}

\begin{corollary}
	Every \(2\)-edge-coloured graph whose underlying graph has tree-width \(2\) admits an improper homomorphism to a monochromatic reflexive copy of \(K_2\).
\end{corollary}

Theorems analogous to Theorems  \ref{thm:ReverseArc} and \ref{thm:AddArc} hold for \(2\)-edge-coloured graphs.
For the analogue of Theorem \ref{thm:ReverseArc} the notion of reversing the orientation of an arc is replaced with the notion of changing the colour of an edge.
For the analogue of Theorem \ref{thm:AddArc} we replace \(2\)-dipath with alternating \(2\)-path.

\section{Complexity of Finding Orientations and \(2\)-edge-colourings that Admit No Improper Homomorphism}\label{sec:Complexity}

In this section we consider the problem of deciding if a graph admits an orientation or a \(2\)-edge-colouring that admits no improper homomorphisms.
As with our work in Section \ref{sec:Structure} we study the problem through the lens of convexity.
For this end we define the following decision problems.\\

\noindent\textbf{Problem:} \emph{COMPLETE CONVEX  2EC}\\
\textbf{Instance:} A graph \(G\).\\
\textbf{Decide:} Does there exist a complete-convex \(2\)-edge-coloured graph \((G,c_G)\)?\\

\noindent\textbf{Problem:} \emph{COMPLETE CONVEX ORIENT}\\
\textbf{Instance:} A graph \(G\).\\
\textbf{Decide:} Does there exist a complete-convex orientation of \(G\)?\\

Theorems \ref{thm:2treeOrient} and \ref{thm:2tree2ec} imply that there are YES instances of \emph{COMPLETE CONVEX ORIENT} that are NO instances of \emph{COMPLETE CONVEX 2EC}. 
These decision problems are not equivalent.
We classify these problems through a reduction from monotone not-all-equal satisfiability.\\

\noindent \textbf{Problem:} \emph{MONOTONE NAE3SAT}\\
\textbf{Instance:} A monotone boolean formula \(Y =(L,F)\) in conjunctive normal form with three variables in each clause.\\
\textbf{Decide:} Does there exist a not-all-equal satisfying assignment for the elements of \(L\)?\\

Without loss of generality, we assume that in an instance of \emph{MONOTONE NAE3SAT} no partition of \(L\) induces a partition of \(F\).
\begin{theorem}\label{thm:NAENPC} \cite{S78}
	The decision problem MONOTONE NAE3SAT is NP-complete.
\end{theorem}

\begin{figure}[h]
	\begin{center}
		\includegraphics[width=.5\linewidth]{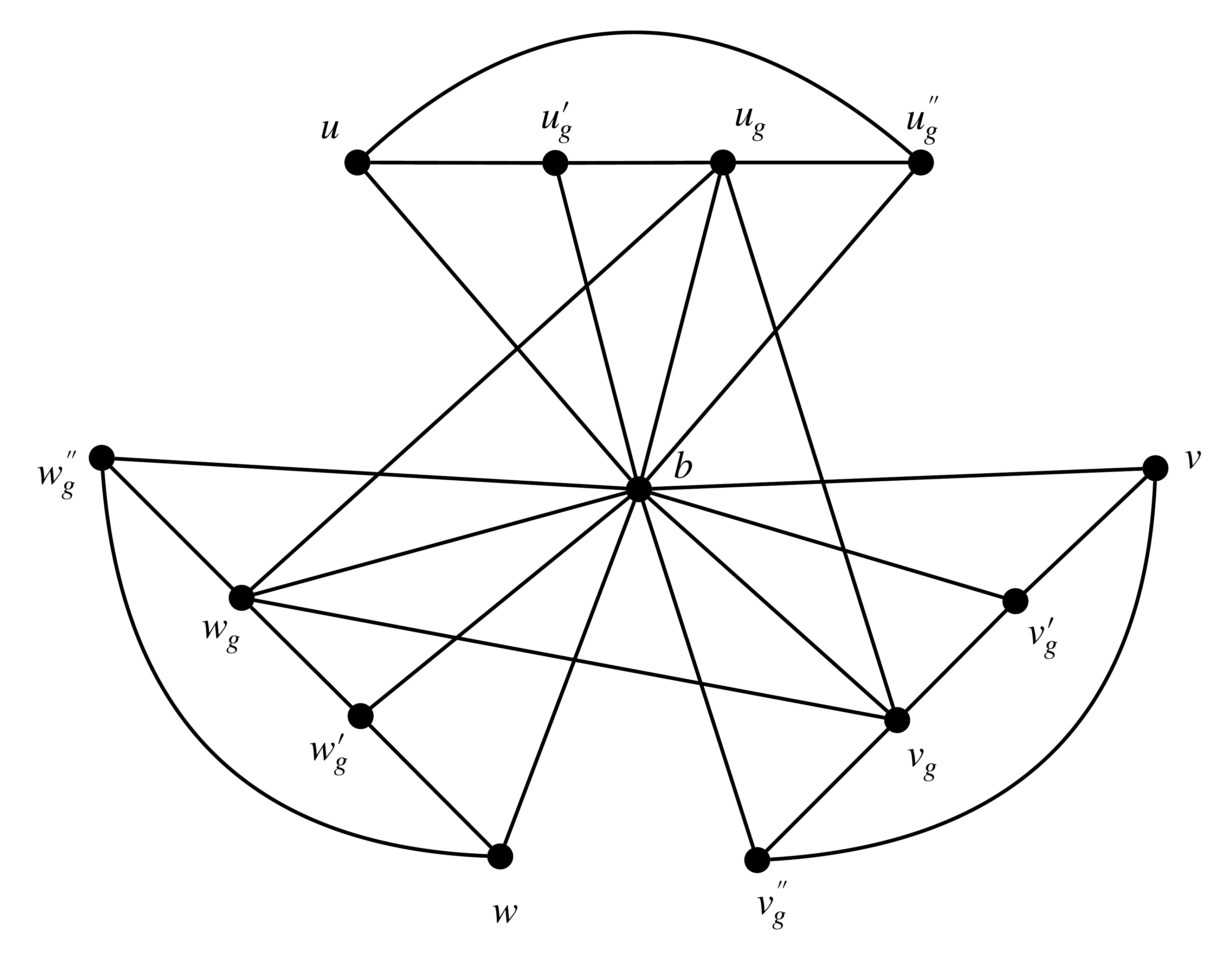}
	\end{center}
	\caption{The clause graph, \(F_g\), for \(g = u \vee v \vee w\).}
	\label{fig:Ff}
\end{figure}

Let \(Y =(L,F)\) be an instance of  \emph{MONOTONE NAE3SAT}.
We construct the graph \(G_Y\) as follows.
For each \(g \in F\) we construct \(F_g\) as shown in Figure \ref{fig:Ff}.
We call these refer to these graphs as \emph{clause graphs}.
We construct \(G_Y\) from the set of clause graphs for \(Y\) as follows:
\begin{itemize}
	\item Identify all vertices labelled \(b\); and
	\item for each \(x \in L\) identify all vertices labelled \(x\).
\end{itemize}
Given \(Y\), the graph \(G_Y\) can be constructed in polynomial time.

We show that there exists complete-convex \(2\)-edge-coloured graph \((G_Y,c_{G_Y})\) if and only if \(Y\) is not-all-equal satisfiable.
For each \(x \in L\), the colour of the edge \(xb\) will represent the assignment for \(x\) in a not-all-equal satisfying assignment for \(Y\).
We begin with three technical lemmas.

\begin{lemma}\label{lem:ecConsistent}
	Consider a complete-convex \(2\)-edge-coloured graph \((G_Y,c_{G_Y})\). For every \(x \in L\) and every \(f \in F\) so that \(x\) is a literal of \(f\) we have \(c_{G_Y}(xb) = c_{G_Y}(x_fb)\).
\end{lemma} 

\begin{proof}
	Assume  \((G_Y,c_{G_Y})\) is complete convex.
	Without loss of generality, assume \(c_{G_Y}(xb) = R\).
	There is only one \(2\)-path with ends \(x\) and \(x_f^\prime\),  the path \(x,b,x_f^\prime\).
	Since \(c_{G_Y}(xb) = R\) and \((G_Y,c_{G_Y})\) is complete convex, it follows that \(c_{G_Y}(x^\prime_fb) = B\).
	By observing that there is only one \(2\)-path with ends \(x_f\) and \(x_f^\prime\) it similarly follows that  \(c_{G_Y}(x_fb) = R\).
\end{proof} 

\begin{lemma}\label{lem:ecNAE}
	Consider complete convex \((G_Y,c_{G_Y})\). For every \(g= u \vee v \vee w\) we have \[\{c_{G_Y}(u_gb),c_{G_Y}(v_gb),c_{G_Y}(w_gb)\} = \{R,B\}.\]
\end{lemma} 

\begin{proof}
	We proceed by contradiction.
	Without loss of generality, assume \(c_{G_Y}(u_gb)=c_{G_Y}(v_gb)=c_{G_Y}(w_gb)  = R\).
	Two  edges of the triangle induced by vertices \(u_g,v_g\) and \(w_g\) must get the same colour, say \(c(u_gv_g) = c(w_gv_g)\).
	However now we notice \(conv\{u_gw_g\} = \{u_g,w_g\}\).
	This is a contradiction as \((G_Y,c_{G_Y})\) is complete convex.
\end{proof}

For a clause \(f\) and a variable \(x\) that appears as a literal of \(f\), denote by \(F_{f,x}\) the subgraph induced by \(\{b,x,x_f,x_f^\prime, x_f^\dprime\}\).
For a clause \(g = u \vee v \vee w\) denote by \(C_g\) the subgraph induced by \(\{b,u_g,v_g,w_g\}\).

\begin{lemma}\label{lem:ECSpreadLemma}
	Consider \((G_Y,c_{G_Y})\). If each of the \(2\)-edge-coloured subgraphs of the form \(F_{f,x}\) and \(C_g\) is complete convex, then \((G_Y,c_{G_Y})\) is complete convex.
\end{lemma}

\begin{proof}
	We proceed by induction on the number of clauses in \(Y\).
	If \(Y\) has a single clause, then \(G_Y = F_f\), where \(f\) is the lone clause in \(F\).
	By Lemma \ref{lem:2ecId},  it follows that \(G_Y\) is complete convex.
	
	Consider now \(|F| = k>1\) and \(f \in F\) with \(f = x \vee y \vee z\).
	Without loss of generality,  \(x\) appears in some other clause of \(F\).
	(Recall that we may assume that no partition of \(L\) induces a partition of \(F\)).
	Let \(Y_f\) be the instance of NAE3SAT formed by removing \(f\) from \(F\).
	By induction \((G_{Y_f},c_{G_{Y_f}})\) is complete convex.
	The result now follows from Lemma \ref{lem:2ecId}.
\end{proof}

For a clause \(f\in F\) that contains literal \(x\), the colour of the edge \(x_fb\) will correspond to the value of the literal \(x\) in the clause \(f\).
Lemma \ref{lem:ecConsistent} implies that for a fixed literal, all such edges have the same colour.
Lemma \ref{lem:ecNAE} implies that for a fixed clause the literals are not all equal.

\begin{lemma}\label{lem:2ecHard}
	Given \(Y  =(L,F)\), an instance of  \emph{MONOTONE NAE3SAT}, there exists a complete-convex \(2\)-edge-coloured graph \((G_Y,c_{G_Y})\) if and only if \(Y\) is not-all-equal satisfiable.
\end{lemma}

\begin{figure}[h]
	\begin{center}
		\includegraphics[width=0.25\linewidth]{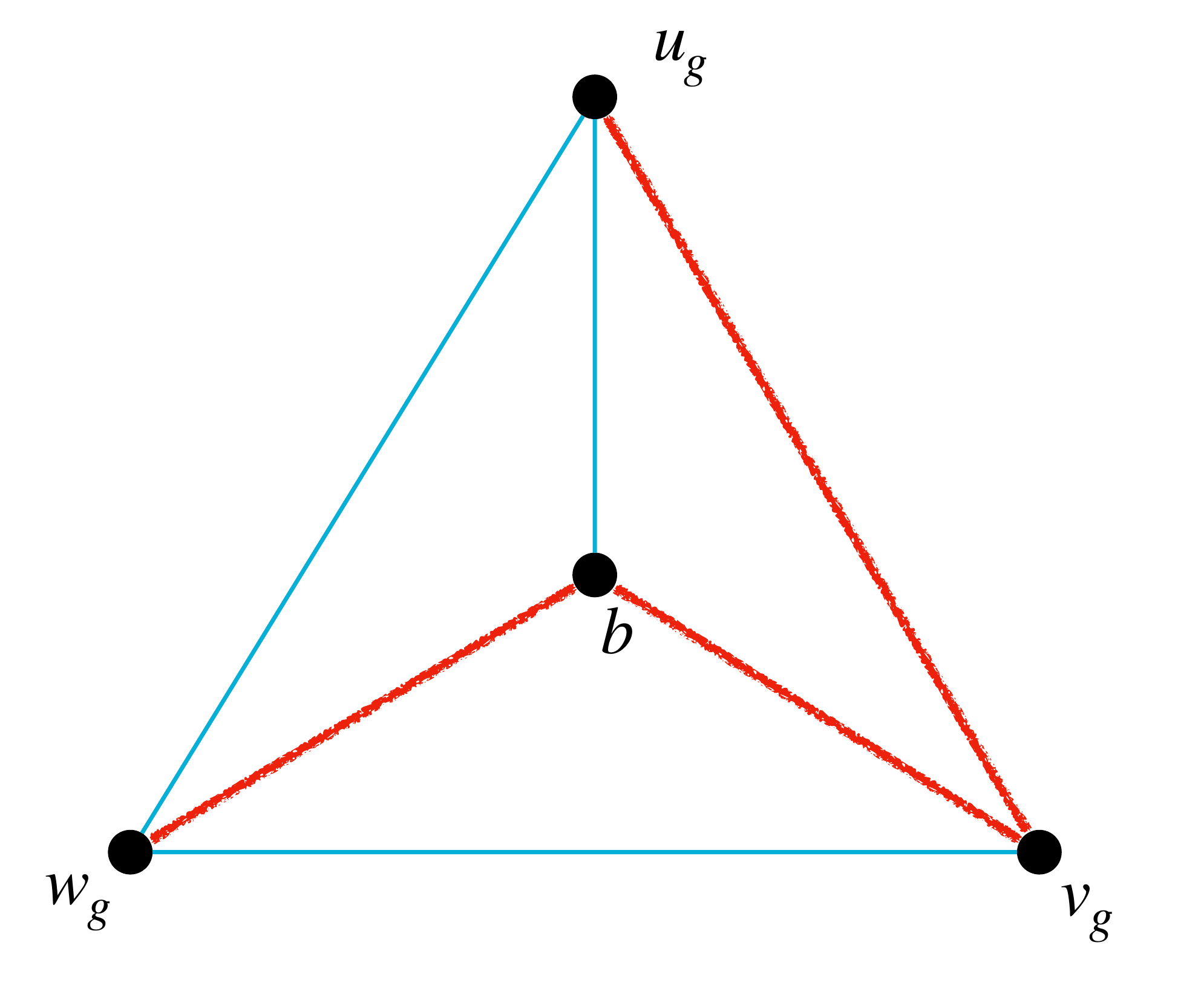}
	\end{center}
	\caption{\(C_g\) when \(s(u)=0\) and \(s(v)=s(w)=1\)}
	\label{fig:SampleCgEC}
\end{figure}

\begin{figure}
	\begin{center}		
	\subfigure[ \(s(x) = 0\)]{\includegraphics[width=0.25\linewidth]{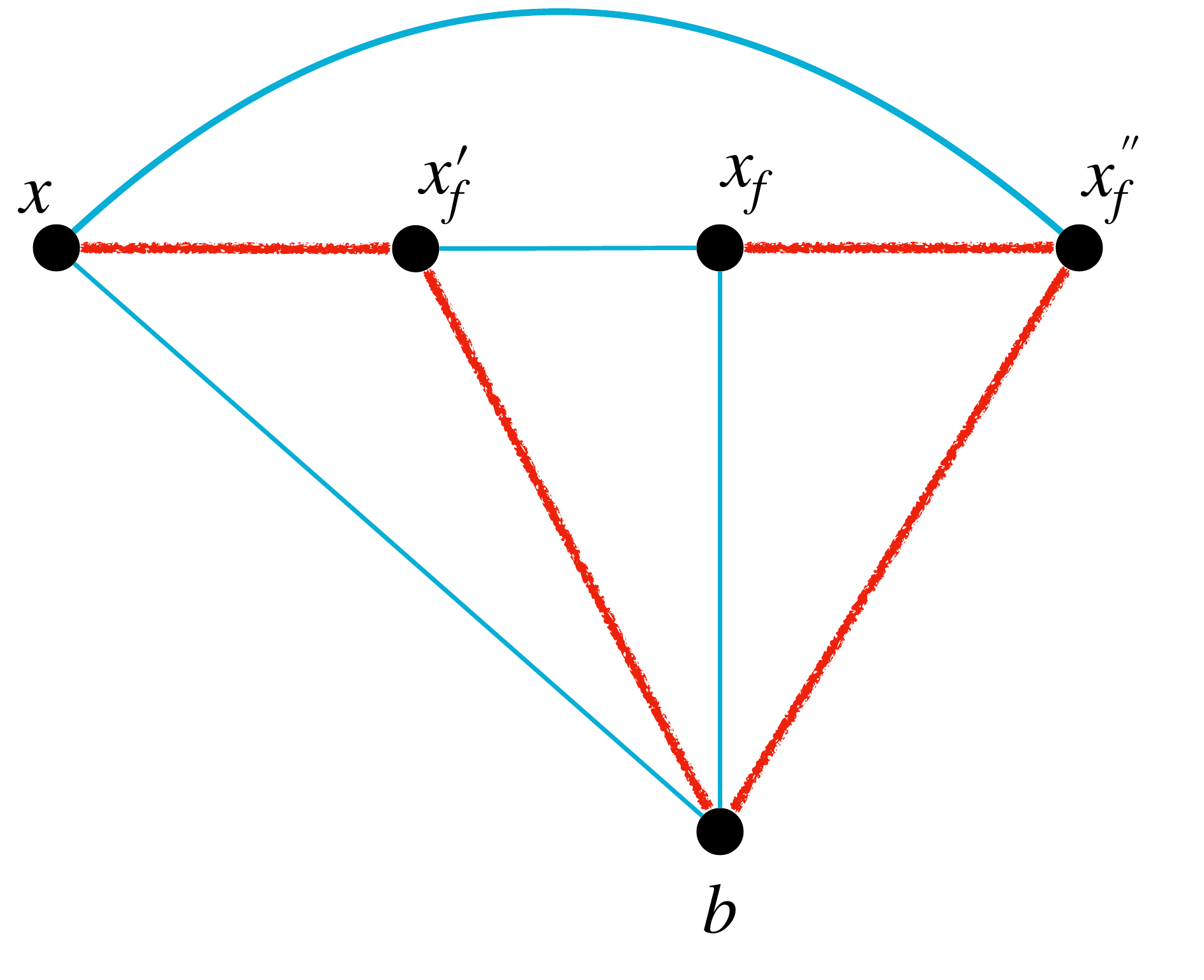}}
	\subfigure[\(s(x) = 1\)]{\includegraphics[width=0.25\linewidth]{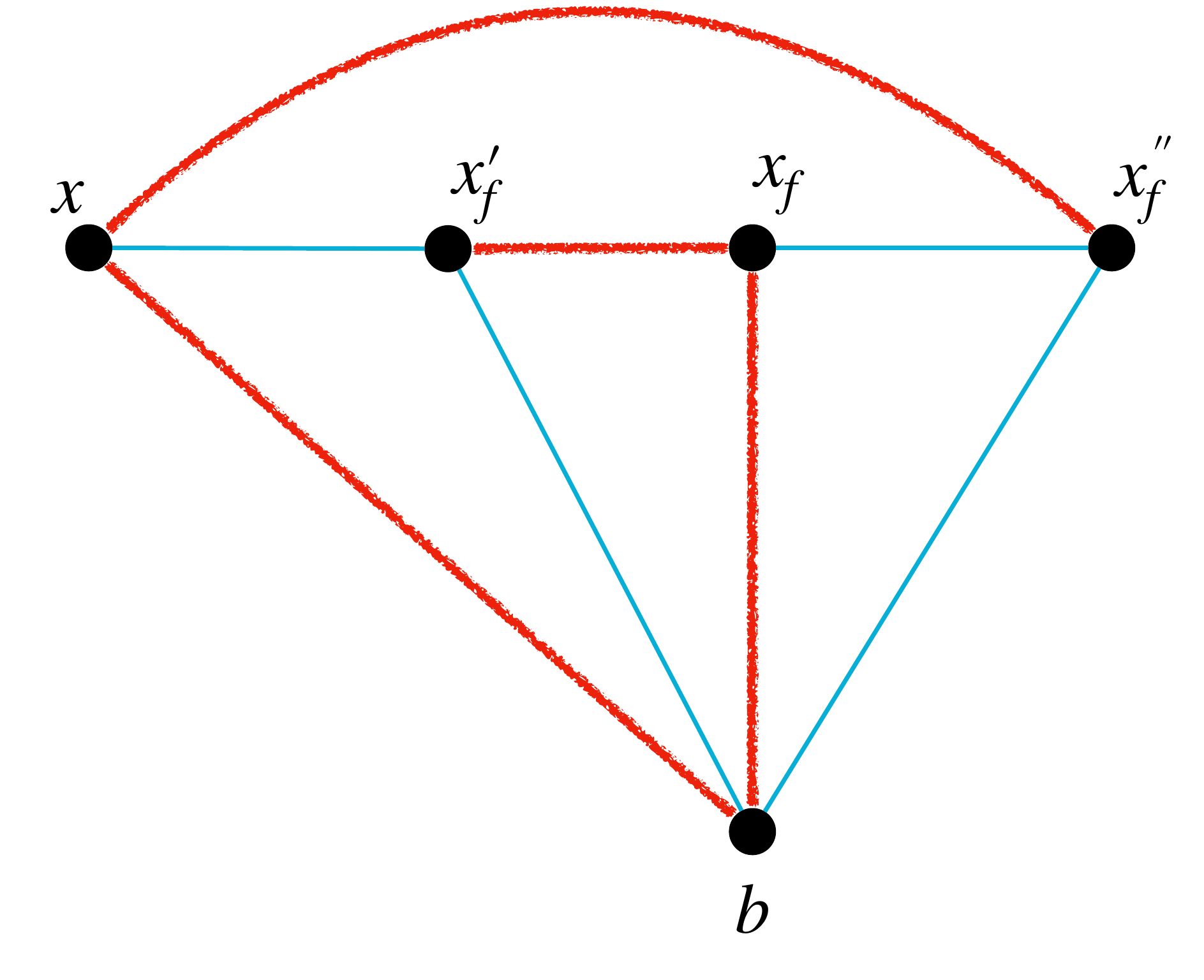}}
	\end{center}
	\caption{\({F_{f,x}}\) for \(s(x) = 0\) and \(s(x) = 1\)}
	\label{fig:SampleFfxEC}
\end{figure}

\begin{proof}
	Let \(Y  =(L,F)\) be an instance of  \emph{MONOTONE NAE3SAT}.
	Consider \((G_Y,c_{G_Y})\), a \(2\)-edge-coloured complete convex graph.
	Construct \(t: L \to \{0,1\}\) so 
	\begin{itemize}
		\item \(t(x) = 1\) when \(c_{G_Y}(xb) = R\); and 
		\item \(t(x) = 0\) when \(c_{G_Y}(xb) = B\).
	\end{itemize}
	
	We claim \(t\) is not-all-equal satisfying for \(Y\).
	Consider \(g \in F\) with \(g = u \vee v \vee w\).
	By Lemma \ref{lem:ecNAE} we have \(\{c_{G_Y}(u_gb),c_{G_Y}(v_gb),c_{G_Y}(w_gb)\} = \{R,B\}\).
	By Lemma \ref{lem:ecConsistent} we have \(c_{G_Y}(u_gb) = c_{G_Y}(ub), c_{G_Y}(v_gb) = c_{G_Y}(vb), c_{G_Y}(w_gb) = c_{G_Y}(wb)\).
	Therefore \(\{t(u), t(v), t(w)\} = \{0,1\}\).
	Therefore \(t\) is not-all-equal satisfying for \(Y\).
	
	Consider now \(s: L \to \{0,1\}\) so that \(s\) is not-all-equal satisfying for \(Y\).
	We construct \(c_{G_Y}\) as follows.
	For all  \(f \in F\) so that \(x\) is a literal of \(f\) and \(s(x) = 1\) let 
	\begin{itemize}
		\item \(c_{G_Y}(xb) = c_{G_Y}(x_fb) = c_{G_Y}(xx_f^\dprime) = c_{G_Y}(x_fx_f^\prime) = R\); and
		\item \(c_{G_Y}(x^\prime_fb) = c_{G_Y}(x_f^\prime x) = c_{G_Y}(x_f^\dprime b) = c_{G_Y}(x_fx_f^\dprime) = B\).
	\end{itemize}
	For all  \(f \in F\) so that \(x\) is a literal of \(f\) and \(s(x) = 0\) let 
	\begin{itemize}
		\item \(c_{G_Y}(xb) = c_{G_Y}(x_fb) = c_{G_Y}(xx_f^\dprime) = c_{G_Y}(x_fx_f^\prime) = B\); and
		\item \(c_{G_Y}(x^\prime_fb) = c_{G_Y}(x_f^\prime x) = c_{G_Y}(x_f^\dprime b) = c_{G_Y}(x_fx_f^\dprime) = R\).
	\end{itemize}
	
	For \(g \in F\) with \(g = u \vee v \vee w\) so that \(s(u) = 0\), \(s(v) = s(w) = 1\) let 
	\begin{itemize}
		\item \(c_{G_Y}(u_gw_g)= c_{G_Y}(w_gv_g) = B\); and
		\item \(c_{G_Y}(u_gv_g)= R\).
	\end{itemize}
	
	For \(g \in F\) with \(g = u \vee v \vee w\) so that \(s(u) = 1\), \(s(v) = s(w) = 0\) let 
	\begin{itemize}
		\item \(c_{G_Y}(u_gw_g)= c_{G_Y}(w_gv_g) = R\); and
		\item \(c_{G_Y}(u_gv_g)= B\). 
	\end{itemize}
	
	(See Figure  \ref{fig:ExampleFTTec} for the case \(s(u)=0\), \(s(v)=s(w)=1\).)
	
	\begin{figure}
		\begin{center}
			\includegraphics[width=0.5\linewidth]{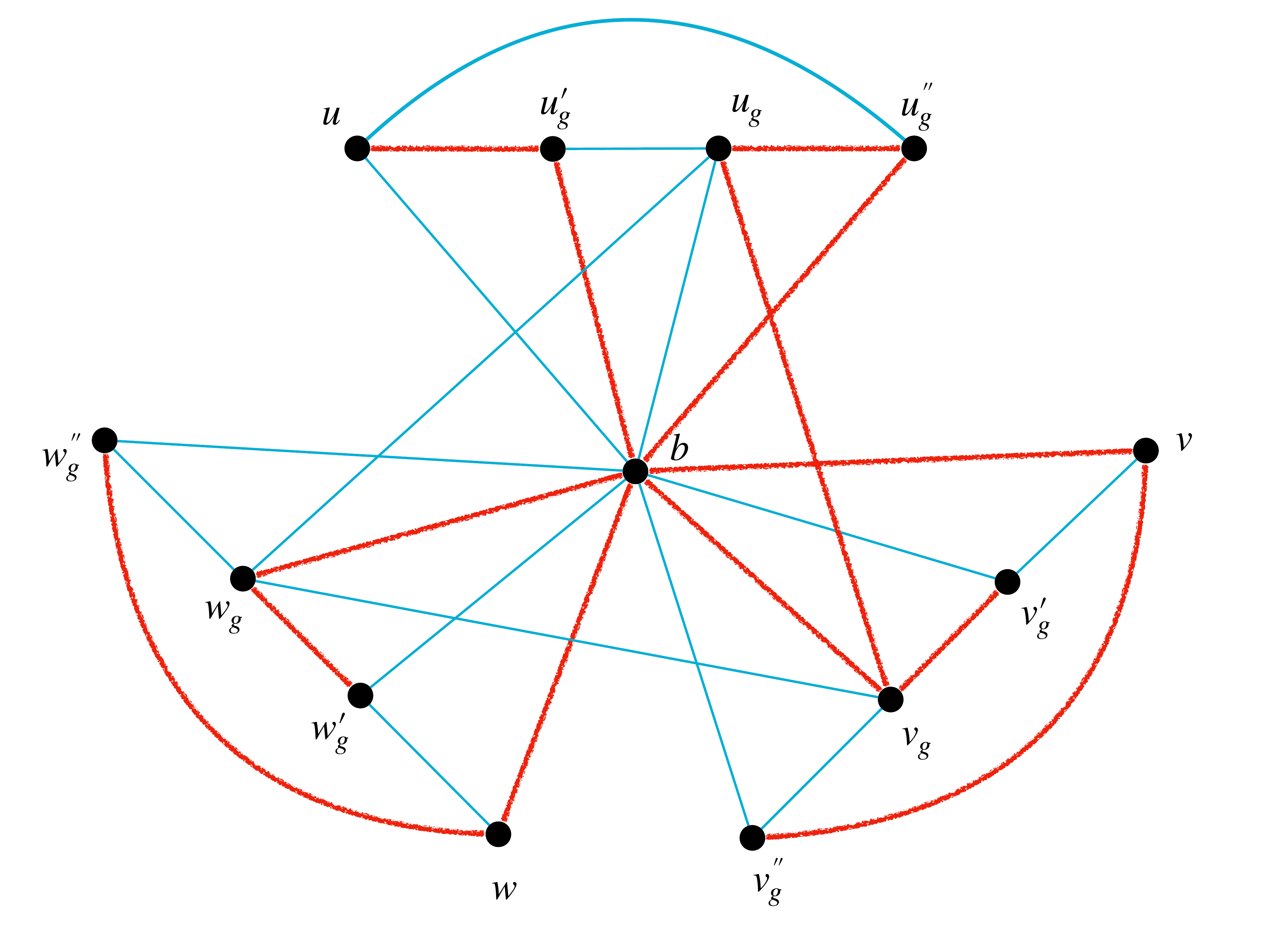}
		\end{center}
		\caption{A \(2\)-edge-coloured clause graph for \(g = u \vee v \vee w\), when \(s(u)=0\) and \(s(v)=s(w)=1\).}
		\label{fig:ExampleFTTec}
	\end{figure}
	
	By inspection, each of the \(2\)-edge-coloured subgraphs of the form \(F_{f,x}\) and \(C_g\) is complete convex (See Figures \ref{fig:SampleCgEC} and \ref{fig:SampleFfxEC})
	Thus by Lemma \ref{lem:ECSpreadLemma} we have that \((G_Y,c_{G_Y})\) is complete convex.
\end{proof}

\begin{theorem}\label{thm:CC2ecNPC}
	The decision problem \emph{COMPLETE CONVEX 2EC} is NP-complete.
\end{theorem}

\begin{proof}
	Verifying that  a \(2\)-edge-coloured graph is complete convex is Polynomial.
	Therefore \emph{COMPLETE CONVEX 2EC} is in NP.
	Given \(Y\), the graph \(G_Y\) can constructed in polynomial time.
	By Lemma \ref{lem:2ecHard} and Theorem \ref{thm:NAENPC}, \emph{COMPLETE CONVEX 2EC} is NP-complete.
\end{proof}

\begin{corollary}
	It is NP-complete to decide if a graph admits a \(2\)-edge-colouring that admits no improper homomorphism.
\end{corollary}

We turn now to oriented graphs.
We proceed similarly as in the argument for \(2\)-edge-coloured graphs.
Given \(Y = (L,F)\) we construct \(G_Y\) as above.
We form \(G_Y^\prime\) from \(G_Y\) by adding a vertex \(g\) and edges \(gu_g, gv_g,gw_g\) and removing vertices \(u^\dprime_g, v^\dprime_g,w^\dprime_g\) for each clause \(g = u \vee v \vee w\).
We show  there exists a complete convex orientation of \(G_Y^\prime\) if and only if \(Y\) is not-all-equal satisfiable.
For each \(x \in L\) the orientation of the edge \(xb\) will represent the assignment for \(x\) in a not-all-equal satisfying assignment for \(Y\).
We begin with three technical lemmas in the spirit of Lemmas \ref{lem:ecConsistent}, \ref{lem:ecNAE} and \ref{lem:ECSpreadLemma}.
The proofs of these lemmas follow similarly to those of Lemmas \ref{lem:ecConsistent}, \ref{lem:ecNAE} and \ref{lem:ECSpreadLemma} and are thus omitted.

\begin{lemma}\label{lem:orientConsistent}
	Consider \(\dir{G^\prime_Y}\) a complete convex orientation of \(G^\prime_Y\). For every \(x \in L\) and every \(f \in F\) so that \(x\) is a literal of \(f\) we have that \(x,b,x_f\) is not a \(2\)-dipath.
\end{lemma} 

\begin{lemma}\label{lem:orientNAE}
	Consider \(\dir{G^\prime_Y}\) a complete convex orientation of \(G^\prime_Y\). For every \(g = u \vee v \vee w\) we have that \(b\) is not a source or sink in the subgraph induced by \(\{u_g,v_g,w_g,b\}\).
\end{lemma} 

For a clause \(f\) and a variable \(x\) that appears as a literal of \(f\), denote by \(F^\prime_{f,x}\) the subgraph induced by \(\{b,x,x_f,x_f^\prime\}\).
For a clause \(g = u \vee v \vee w\) denote by \(C^\prime_g\) the subgraph induced by \(\{g,b,u_g,v_g,w_g\}\).

\begin{lemma}\label{lem:OrientSpreadLemma}
	Consider \(\dir{G^\prime_Y}\). If each of the oriented subgraphs of the form \(\dir{F^\prime_{f,x}}\) and \(\dir{C^\prime_g}\) is complete convex, then \(\dir{G^\prime_Y}\) is complete convex.
\end{lemma}

\begin{lemma}\label{lem:orientHard}
	Given \(Y  =(L,F)\), an instance of \emph{MONOTONE NAE3SAT}, there exists a complete convex orientation \(\dir{G^\prime_Y}\) if and only if \(Y\) is not-all-equal satisfiable.
\end{lemma}

\begin{proof}
	Let \(Y  =(L,F)\) be an instance of  \emph{MONOTONE NAE3SAT}.
	
	Consider \(\dir{G^\prime_Y}\), an orientation of \(G^\prime_Y\).
	Construct \(t: L \to \{0,1\}\) so 
	\begin{itemize}
		\item \(t(x) = 0\) when \(xb \in A_{\dir{G^\prime_Y}}\) and
		\item \(t(x) = 1\) when \(bx \in A_{\dir{G^\prime_Y}}\).
	\end{itemize}
	
	We claim \(t\) is not-all-equal satisfying for \(Y\).
	Consider \(g \in F\) with \(g = u \vee v \vee w\).
	By Lemma \ref{lem:orientNAE} we have that \(b\) is not a source or a sink in the subgraph induced by \(\{u_g,v_g,w_g,b,g\}\).
	Therefore \(\{t(u), t(v), t(w)\} = \{0,1\}\).
	By Lemma \ref{lem:orientConsistent} we have that the edge \(ub\) is oriented towards \(b\) if and only if the edge \(u_gb\) is oriented towards \(b\).
	Similarly, the edge \(vb\) is oriented towards \(b\) if and only if the edge \(v_gb\) is oriented towards \(b\) and the edge \(wb\) is oriented towards \(b\) if and only if the edge \(w_gb\) is oriented towards \(b\).
	Therefore \(t\) is not-all-equal satisfying for \(Y\).	
	
	Consider now \(s: L \to \{0,1\}\) so that \(s\) is not-all-equal satisfying for \(Y\).
	For all \(f \in F\) so that \(x\) is a literal of \(f\) and \(s(x) = 1\) 
	\begin{itemize}
		\item orient the edge \(xx_f^\prime\) to have its tail at \(x\);
		\item orient the edge  \(x_fx_f^\prime\) to have its tail at \(x_f\);
		\item orient edges \(xb\) and \(x_fb\) to have their tails at \(b\); and
		\item orient the edge \(x_f^\prime b\) to have its head at \(b\).
	\end{itemize}
	For all  \(f \in F\) so that \(x\) is a literal of \(f\) and \(s(x) = 0\) 
	\begin{itemize}
		\item orient the edge  \(xx_f^\prime\) to have its head at \(x\);
		\item orient the edge  \(x_fx_f^\prime\) to have its head at \(x_f\);
		\item orient edges \(xb\) and \(x_fb\) to have their heads at \(b\); and
		\item orient the edge \(x_f^\prime b\) to have its tail at \(b\).
	\end{itemize}
	
	For \(g \in F\) with \(g = u \vee v \vee w\) so that \(s(u) = 0\), \(s(v) = s(w) = 1\)
	\begin{itemize}
		\item orient the edge \(v_gw_g\) to have its tail at \(w_g\);
		\item orient the edge \(u_gg\) to have its tail at \(u_g\); and
		\item orient \(u_g,v_g,g,w_g\) as a directed \(4\)-cycle.
	\end{itemize}
	
	For \(g \in F\) with \(g = u \vee v \vee w\) so that \(s(u) = 1\), \(s(v) = s(w) = 0\)
	\begin{itemize}
		\item orient the edge \(v_gw_g\) to have its head at \(w_g\); 
		\item orient the edge \(u_gg\) to have its head at \(u_g\); and
		\item orient \(w_g,g,v_g,u_g\) as a directed \(4\)-cycle.
	\end{itemize}
	
	(See Figure  \ref{fig:ExampleTFForient} for the case \(s(u)=1\), \(s(v)=s(w)=0\).)
	
	\begin{figure}
		\begin{center}
			\includegraphics[width=0.5\linewidth]{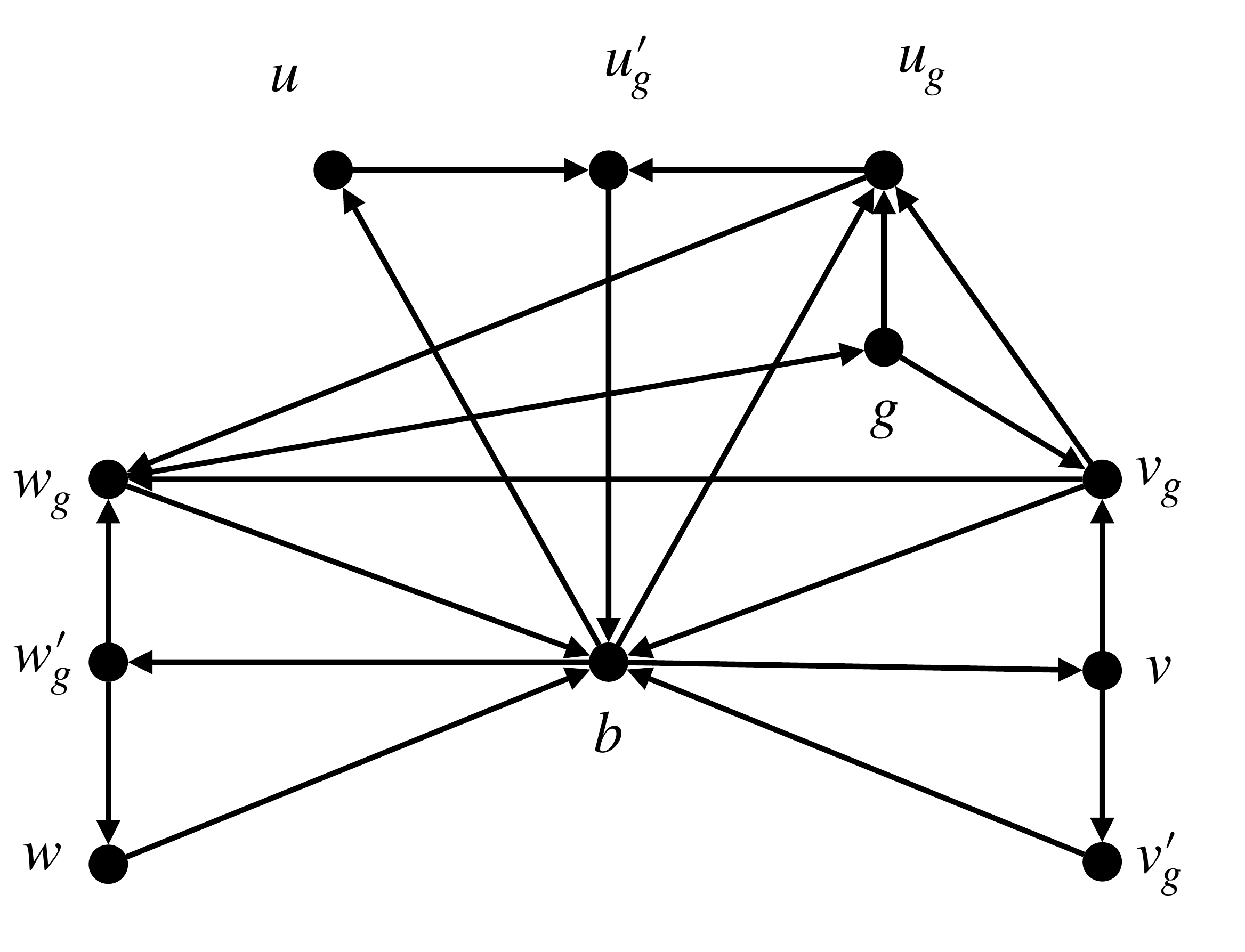}
		\end{center}
		\caption{An oriented clause graph for \(g = u \vee v \vee w\), when \(s(u)=1\) and \(s(v)=s(w)=0\).}
		\label{fig:ExampleTFForient}
	\end{figure}
	
	For \(g = u \vee v \vee w\), we observe that the oriented graph \(\dir{C^\prime_g}\) is complete convex (See Figure \ref{fig:SampleCgOrient}). 
	For \(x\in L\) and \(f \in F\) so that \(x\) is a literal of \(f\), we observe that the oriented graph \(\dir{F^\prime_{f,x}}\) is complete convex (See Figure \ref{fig:SampleFfxOrient}).
\begin{figure}
	\begin{center}
		\includegraphics[width=0.5\linewidth]{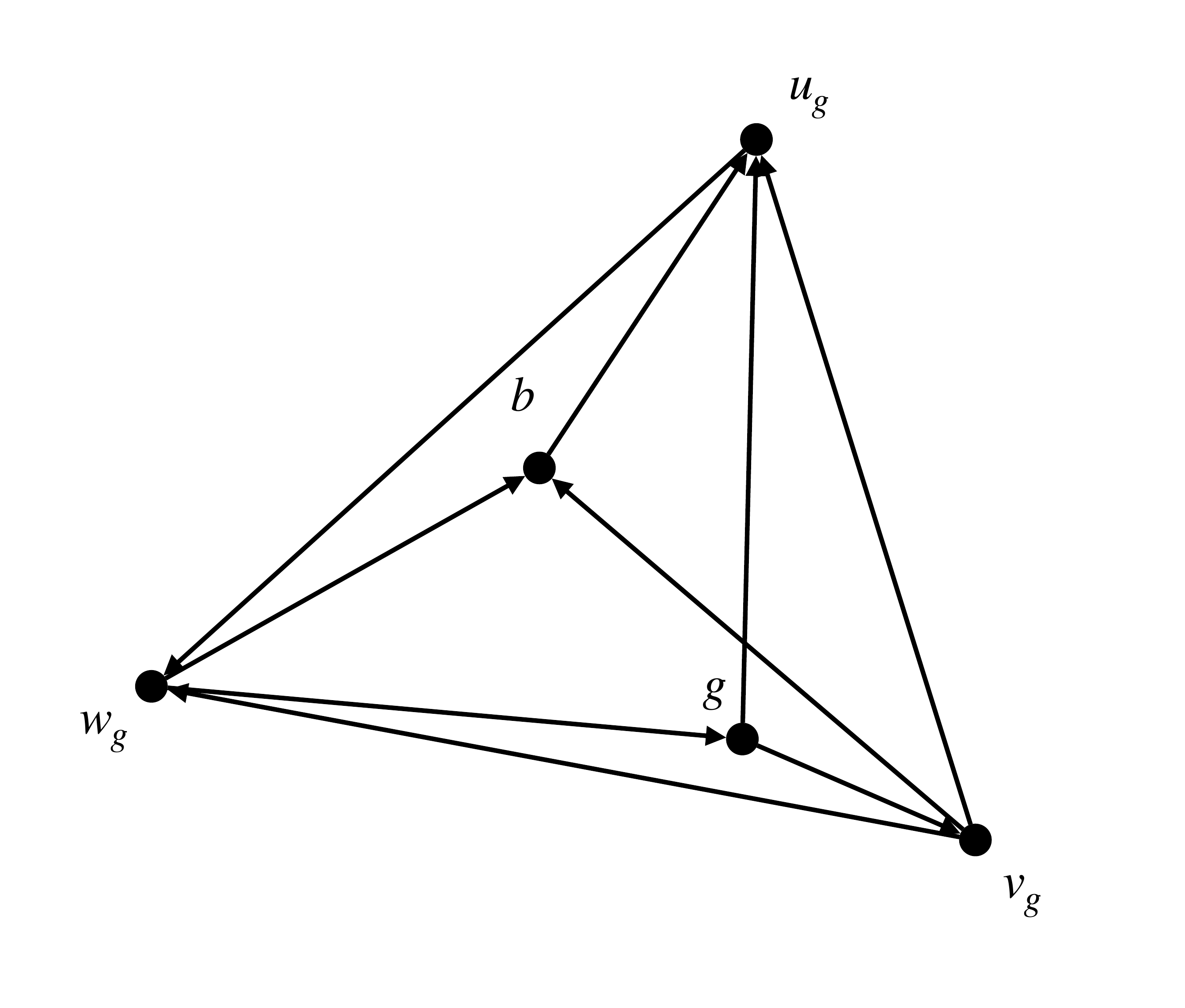}
	\end{center}
	\caption{\(C^\prime_g\) for the case \(s(u)=1\) and \(s(v)=s(w)=0\)}
	\label{fig:SampleCgOrient}
\end{figure}

\begin{figure}
	\begin{center}		
		\subfigure[ \(s(x) = 0\)]{\includegraphics[width=0.25\linewidth]{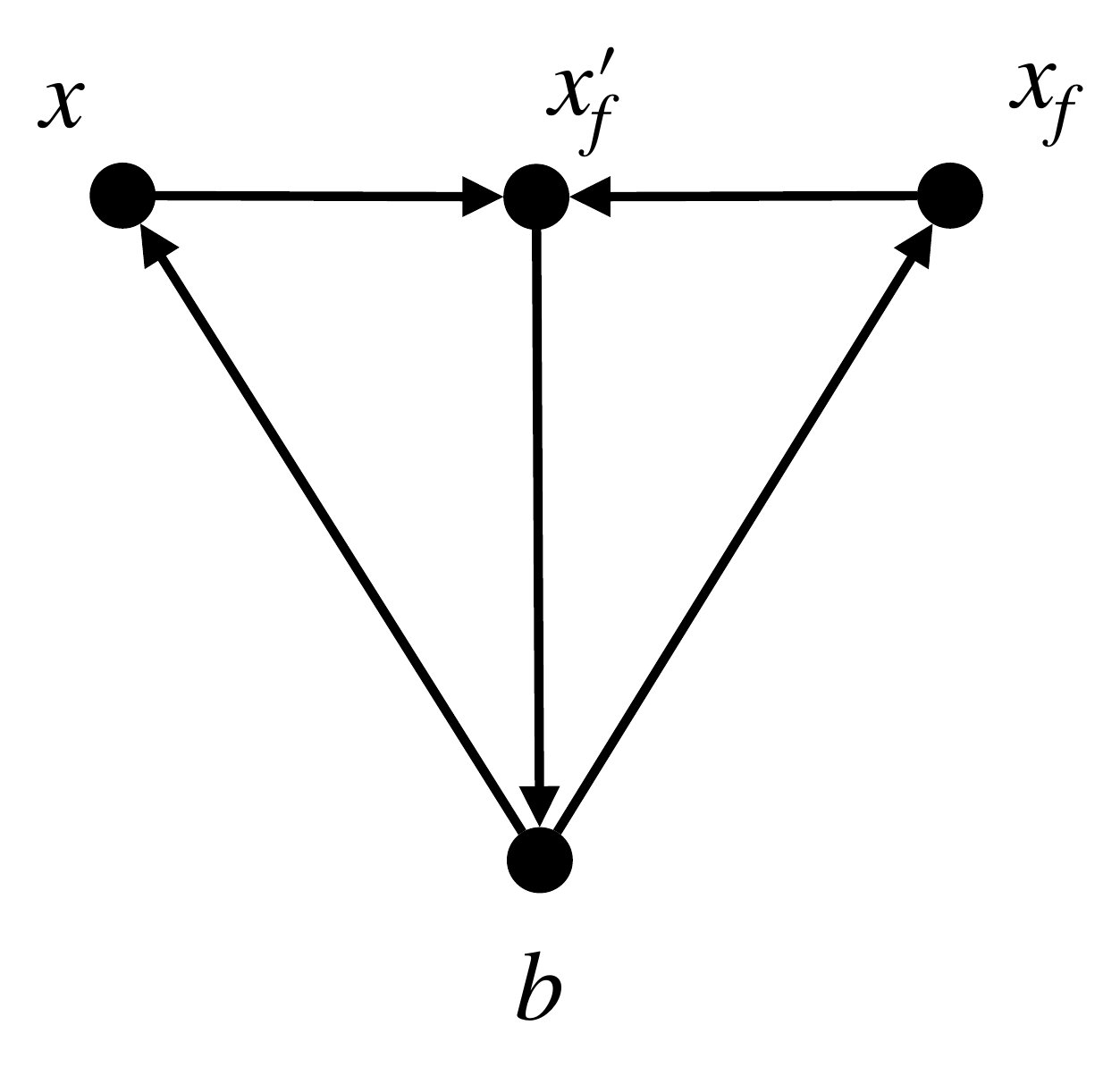}}
		\subfigure[\(s(x) = 1\)]{\includegraphics[width=0.25\linewidth]{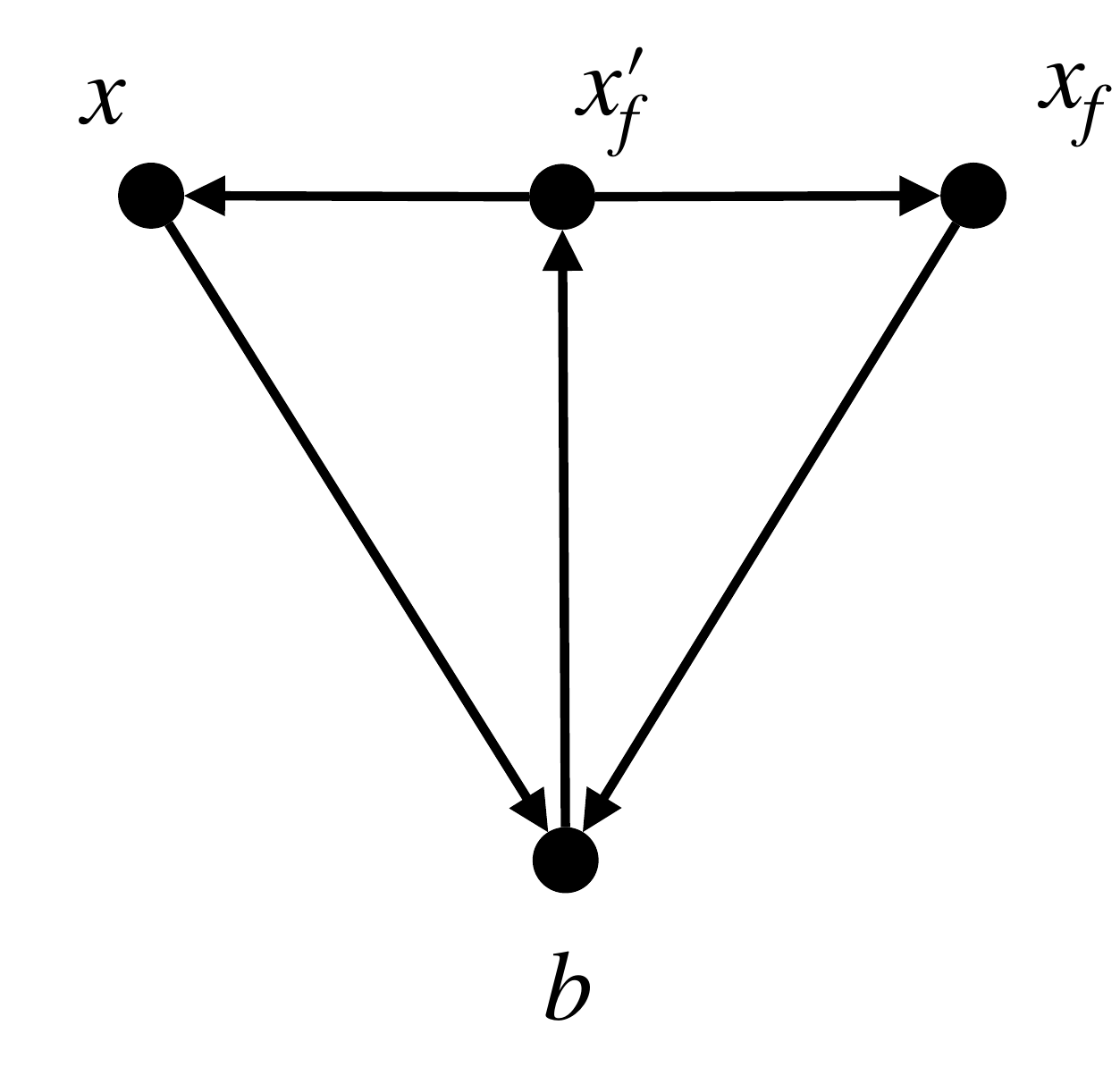}}
	\end{center}
	\caption{\({F^\prime_{f,x}}\) for \(s(x) = 0\) and \(s(x) = 1\)}
	\label{fig:SampleFfxOrient}
\end{figure}

	The result now follows from Lemma \ref{lem:OrientSpreadLemma}.
\end{proof}
\begin{theorem}
	The decision problem \emph{COMPLETE CONVEX ORIENT} is NP-complete.
\end{theorem}

\begin{corollary}
	It is NP-complete to decide if a graph admits an orientation that admits no improper homomorphism.
\end{corollary}

\cite{DP18} show each of \emph{COMPLETE CONVEX EC} and \emph{COMPLETE CONVEX ORIENT} are polynomial when restricted to complete graphs 
Theorems \ref{thm:2treeOrient} and \ref{thm:2tree2ec}  imply that each of \emph{COMPLETE CONVEX EC} and \emph{COMPLETE CONVEX ORIENT} are polynomial when restricted to graphs with tree-width \(2\).

\section{The Chromatic Number of Minor-Closed Families of Oriented Graphs}\label{sec:MinorClosed}
Recall that one may use homomorphism to define a notion of proper vertex colouring for oriented graph that, in some sense, respects the orientation of the arcs.

Let \(\dir{G}\) be an irreflexive oriented graph.
The \emph{chromatic number} of \(\dir{G}\) is the least integer \(k\) that \(\dir{G} \to \dir{T}\) for some tournament \(\dir{T}\) with \(k\) vertices.
We denote this parameter as \(\chi(\dir{G})\).
For a family of irreflexive oriented graphs \(\dir{\mathcal{F}}\) we define \(\chi(\dir{\mathcal{F}})\) to be the least integer \(k\) such that \(\chi(\dir{F}) \leq k\) for each \(\dir{F} \in \dir{\mathcal{F}}\).

This notion of chromaticity was first introduced by \cite{CO94}.
In this work he shows that formulas expressed in the monadic second-order logic of graph can be decided in polynomial time for graphs with bounded tree-width. 
A major landmark in the study of such colourings of is the upper bound on the chromatic number of orientations of planar graphs.

\begin{theorem}\label{thm:Orient80}\cite{RASO94}
	Let \(P\) be an irreflexive planar graph. For any orientation \(\dir{P}\) we have \(\chi (\dir{P}) \leq 80\).
\end{theorem}

The proof of this result does not rely directly on planarity, rather it follows from the fact that every planar graph admits an acyclic \(5\)-colouring (see \cite{Gr83}.)
Thus it remains possible that this bound can be  improved.
This possibility is buttressed by a complete lack of examples of orientations of planar graphs whose chromatic number is between \(19\) and \(80\).
Though significant work has gone into the study of the chromatic number of orientations of planar graphs the last \(25\) years, most meaningful progress has been made on restricted classes of planar graphs, such as those with bounded girth (see \cite{B07,B05,B98,O14,P09}).
A notable exception to this lack of progress for the general case is work by \cite{S02b} (see Theorem \ref{thm:SimpleSmolinkova} below) resulting from the following relaxation of the definition of chromatic number to allow for non-trivial homomorphisms to reflexive targets.

Let \(\dir{G}\) be an oriented graph.
The \emph{simple chromatic number} of \(\dir{G}\) is the least integer \(k\) so that there exists a non-trivial homomorphism \(\dir{G} \to \dir{T}\) for some reflexive tournament \(\dir{T}\) with \(k\) vertices.
We denote this parameter as \(\chi_s(\dir{G})\).
We analogously define \(\chi_s(\dir{\mathcal{F}})\).

\begin{theorem}\label{thm:SimpleSmolinkova}\cite{S02b}
	For \(\dir{\mathcal{P}}\) the family of orientations of irreflexive planar graphs we have \(\chi(\dir{\mathcal{P}}) = \chi_s(\dir{\mathcal{P}}).\)
\end{theorem}

Our work on the study of complete-convex oriented graphs and the definitions above directly imply the following.

\begin{theorem}\label{thm:CCchi}
	Let  \(\dir{G}\) be an irreflexive oriented graph. If \(\dir{G}\) is complete convex, then \(\chi(\dir{G})  = \chi_s(\dir{G})\).
\end{theorem}

With an eye towards the study of orientations of planar graphs, we use this theorem to study the simple chromatic number of minor-closed families of graphs.

\begin{theorem}\label{thm:Contract}
	Let \(\mathcal{F}\) be a family of irreflexive graphs that is closed with respect to edge contraction.
	Let \(\dir{\mathcal{F}}_c\) denote the set of complete-convex orientations of elements of \(\mathcal{F}\).
	We have \(\chi_s({\dir{\mathcal{F}}}) = \chi({\dir{\mathcal{F}}_c})\).
\end{theorem}

\begin{proof}
	Let \(\dir{\mathcal{F}}\) be the set of orientations of some family \(\mathcal{F}\) that is closed with respect to edge contraction.
	Let \(\dir{\mathcal{G}} \subseteq \dir{\mathcal{F}}\) be the family of oriented graphs for which \(\chi(\dir{G}) = \chi(\dir{\mathcal{F}})\) for each \(\dir{G} \in \dir{\mathcal{G}}\).
	Consider \(\dir{H} \in \dir{\mathcal{G}}\) with the fewest number of vertices.
	We claim \(\dir{H}\) is complete convex.
	
	Assume otherwise and consider \(uv \in A_{\dir{H}}\) so that \(conv(uv) \neq V_H\).
	Let \(S = conv(uv)\).
	Let \(\dir{H_S}\) be the oriented graph formed from \(\dir{H}\) by identifying vertices of \(S\) into a single vertex named \(s\).
	By the minimality of \(\dir{H}\) we have \(\chi_s(\dir{H_S}) < \chi_s(\dir{H}) = \chi_s({\dir{\mathcal{F}}})\).
	Let \(\phi\) be a \(\chi_s(\dir{H_S})\)-colouring of \(\dir{H_S}\).
	We construct \(\beta\),  a \(\chi_s(\dir{H_S})\)-colouring of \(\dir{H}\), as follows
	
	\[	\beta(x)=
	\begin{cases}
		\phi(x) & x \notin S\\
		\phi(s) & x \in S
	\end{cases}\]
	
	The existence of \(\beta\) contradicts \(\chi_s(\dir{H}) = \chi_s({\dir{\mathcal{F}}})\).
	Thus \(\dir{H}\) is complete convex.
	
	Let \(\dir{\mathcal{F}_c}\) be the set of complete-convex elements of  \(\dir{\mathcal{F}}\).
	Since \(\dir{H} \in \dir{\mathcal{F}_c}\) we have \(\chi_s(\dir{\mathcal{F}_c}) = \chi_s(\dir{\mathcal{F}})\).
	By Theorem \ref{thm:CCchi} we have \(\chi(\dir{F}) = \chi_s(\dir{F})\) for each  \(\dir{F} \in \dir{\mathcal{F}_c}\).
	Therefore \(\chi(\dir{\mathcal{F}_c}) = \chi_s(\dir{\mathcal{F}_c}) = \chi_s(\dir{\mathcal{F}})\).	
\end{proof}

\begin{corollary}\label{cor:MinorClosed}
	If \(\mathcal{F}\) is a minor-closed family of irreflexive graphs, then \(\chi_s({\dir{\mathcal{F}}}) = \chi({\dir{\mathcal{F}}_c})\) .
	%and \(\chi_s({\ec{\mathcal{F}}}) = \chi({\ec{\mathcal{F}}_c})\).
\end{corollary}

By way of demonstration, we apply Corollary \ref{cor:MinorClosed} to some well-known families of minor-closed irreflexive graphs.
\begin{theorem}
	\noindent
	\begin{enumerate}
		\item For \(\mathcal{W}\) the family irreflexive graphs with maximum degree \(2\), we have \(\chi_s({\dir{\mathcal{W}}}) = 3\).
		\item For \(\mathcal{G}\) the family of forests, we have  \(\chi_s({\dir{\mathcal{G}}}) = 2\).
		\item For \(\mathcal{R}\) the family of irreflexive graphs with tree-width at most \(2\), we have \(\chi_s({\dir{\mathcal{R}}}) = 3\).
	\end{enumerate}
\end{theorem}

\begin{proof}
	\noindent
	\begin{enumerate}
		\item 
		The only  elements  of \(\dir{\mathcal{W}_c}\) are the directed path with two vertices and the directed cycle with three vertices.
		The directed \(3\)-cycle has chromatic number \(3\).
		By Corollary \ref{cor:MinorClosed} it follows that \(\chi_s({\dir{\mathcal{W}}}) = 3\).		
		\item		
		The only  element  of \(\dir{\mathcal{G}_c}\) is the directed path with two vertices.
		By Corollary \ref{cor:MinorClosed} it follows that \(\chi_s({\dir{\mathcal{G}}}) = 2\).		
		\item 		
		By Theorem \ref{thm:2treeOrient}, the only  elements of \(\dir{\mathcal{R}_c}\) are the directed path with two vertices and orientations of \(2\)-trees in which each copy of \(K_3\) is oriented as a directed three cycle.
		It easily checked that each such orientation of a \(2\)-tree admits a proper homomorphism to the directed three cycle.
		Thus \(\chi_s({\dir{\mathcal{R}}}) = 3\).
	\end{enumerate}
\end{proof}

Each of these results appear in previous work   (see \cite{DP18,S02b}).
However, we note that in each case the result is found by finding a universal target for homomorphism  over all graphs in the relevant family.
Here this process is simplified by studying only those complete convex elements of the family.
As such Corollary \ref{cor:MinorClosed} has the potential to be a powerful tool in studying the simple chromatic number of orientations of minor-closed families.
To wit, combining Corollary \ref{cor:MinorClosed} with the statement of Theorem \ref{thm:SimpleSmolinkova} yields the following result.

\begin{corollary}\label{cor:Planar}
	Let \(\dir{\mathcal{P}}\) be the family of orientations of planar graphs.
	We have \(\chi({\dir{\mathcal{P}}}) = \chi({\dir{\mathcal{P}}_c})\).
\end{corollary}

Thus to improve the bound on the chromatic number of orientations of planar graphs given in Theorem \ref{thm:Orient80} one may  restrict attention to those orientations that are complete convex. 
We comment further on this approach in Section \ref{sec:Conclusion}.

\section{Conclusion and Outlook}\label{sec:Conclusion}
The similarity in these results for oriented graphs and \(2\)-edge-coloured graphs herein is expected based on previous work on homomorphisms of these objects. 
Indeed, the respective study of the chromatic number of oriented and \(2\)-edge-coloured graphs is littered with results and techniques that are strikingly similar.

For fixed \((m,n) \neq (0,0)\) an \((m,n)\)-mixed graph is a graph that permits \(m\) different colours of arcs and \(n\) different colours of edges.
In particular,  graphs are \((0,1)\)-mixed graphs, oriented graphs are \((1,0)\)-mixed graphs and \(2\)-edge-coloured graphs are \((0,2)\)-mixed graphs.
The definitions of colouring and homomorphism for graphs, oriented graphs and \(2\)-edge-coloured graph can be generalized using \((m,n)\)-mixed graphs.
It is unsurprising then that definitions, theorems and constructions for oriented and \(2\)-edge-coloured colourings often extend to the more general \((m,n)\)-mixed graph setting.
Theorem  \ref{thm:Orient80} and an analogue of this result for \(2\)-edge-coloured graphs by \cite{AM98} were shown to be special cases of a more general theorem for \((m,n)\)-mixed graphs by \cite{NR00}.
Theorem \ref{thm:SimpleSmolinkova} remains true for planar \(2\)-edge-coloured graphs and \((m,n)\)-mixed graphs (see \cite{D15,DP18}).
As do the results in Section \ref{sec:MinorClosed}.
Similarities also occur in the study of the chromatic number of bounded degree oriented, \(2\)-edge-coloured and \((m,n)\)-mixed graphs  (see \cite{Das17,KSZ97}).
We note however the contrast between the classification of complete convex orientations and \(2\)-edge-coloured \(2\)-trees given in Theorems \ref{thm:2treeOrient} and \ref{thm:2tree2ec}.
Such contrasting results also exist in the study of chromatic polynomials of these objects (see \cite{B20,C19}).
It remains unclear for which properties one can expect similar results for oriented graphs, \(2\)-edge-coloured graphs and \((m,n)\)-mixed graph.

Corollary \ref{cor:Planar} implies that the study of the chromatic number of orientations of planar graphs may be restricted to those that are complete convex.
Further study into complete-convex planar graphs may reveal structure that can be exploited in the construction of universal targets for this family. 
Many results in the study of universal targets for oriented graphs proceed by contradiction via minimum counter-example.  
Lemma \ref{lem:degree2Reduce} and Theorems \ref{thm:ReverseArc} and \ref{thm:AddArc} provide tools to modify complete-convex oriented graphs while maintaining complete convexity.

For an oriented graph \(\dir{G}\) that is not complete convex our work herein implies the existence of a reflexive oriented graph \(\dir{H}\) so that there is an improper homomorphism of \(\dir{G}\) to \(\dir{H}\).
The construction in the proof of Theorem \ref{thm:onlyG} provides a method to construct such an \(\dir{H}\).
However our results do not provide insight into the problem of deciding if an oriented graph admits an improper homomorphism to some fixed reflexive target \(\dir{H}\).
We propose this  as a future area for research.

\section{Acknowledgments}
The authors acknowledge the contribution of the anonymous peer who, in review, provided the statement and proof of Theorem \ref{thm:2n-3} as a generalization of a result in the submitted manuscript.

\bibliographystyle{abbrvnat}
\bibliography{references}

\end{document}